\documentclass[prl,twocolumn,amsmath,amssymb,floatfix,letter]{revtex4-1}
\usepackage[dvips]{graphicx}
\usepackage{color,array,dcolumn}
\usepackage{amsmath}
\usepackage{amssymb}

\begin{document}

\title{Hidden by graphene -- towards effective screening of interface van der Waals interactions via monolayer coating}

\author{Alberto Ambrosetti}
\author{Pier Luigi Silvestrelli}
\affiliation{
Dipartimento di Fisica e Astronomia, Universit\`{a} degli Studi di Padova, via Marzolo 8, \textsl{35131}, Padova, Italy}
\email{ambroset@pd.infn.it}

\begin{abstract}
Recent atomic force microscopy (AFM) experiments~[ACS Nano {\bf 2014}, 8, 12410-12417] conducted on graphene-coated SiO$_2$ 
demonstrated that monolayer graphene (G) can effectively screen dispersion van der Waals (vdW) interactions deriving from the underlying substrate: 
despite the single-atom thickness of G, the AFM tip was almost insensitive to SiO$_2$, and the 
tip-substrate attraction was essentially determined only by G. This G vdW {\it opacity}  has far reaching implications, encompassing 
stabilization of multilayer heterostructures, micromechanical phenomena or even heterogeneous catalysis.
Yet, detailed experimental control and high-end applications of this phenomenon await sound physical understanding of the underlying physical mechanism.
By quantum many-body analysis and ab-initio Density Functional Theory, here we address this challenge providing theoretical rationalization of the observed G vdW {\it opacity} for weakly interacting substrates. The non-local density response and ultra slow decay of the G vdW interaction ensure compensation between standard attractive terms and many-body repulsive contributions, enabling vdW {\it opacity} over a broad range of adsorption distances. vdW {\it opacity} appears most efficient in the low frequency limit and extends beyond London dispersion
including electrostatic Debye forces. By virtue of combined theoretical/experimental validation, G hence emerges as a promising ultrathin {\it shield} 
for modulation and switching of vdW interactions at interfaces and complex nanoscale devices.
\end{abstract}

\maketitle

The advent of graphene~\cite{Graphene-science} and ensuing experimental progress in synthesis and manipulation of complex nanoscale heterostructures~\cite{Graphene-natmat,fernandez}
has undoubtedly set new frontiers in electronics~\cite{Graphene-electronics}, optics~\cite{Graphene-optics}, and functional nanomaterials~\cite{Regueira}. 
Strong intercarbon bonding
and low chemical reactivity of mono-layer graphene (G) largely contributed to this success: 
weakness of chemical interactions at the interface with many metallic and finite-gap substrates~\cite{thygesen,Hu,Fan} or adsorbates~\cite{jpcc2011} 
facilitates isolation of large high-quality G sheets, while non-covalent van der Waals (vdW) forces typically emerge as the leading ~\cite{Noa-PRL-2010} 
stabilization mechanism.

Understanding vdW interactions in G and 2D materials is essential for predictively modeling structural, 
response~\cite{Reilly-JPCL} or even electronic~\cite{Nicola-PRL,Nicola-PRM} properties of growingly complex heterostructures
and interfaces. 
However, recent experiments evidenced pronounced anomalies~\cite{grafsurf,Tautz-NatureComm,Jacobs1,Jacobs2,Khlobystov-ACSNano} of the vdW interaction in low 
dimensional nanoscale materials, hence challenging time honoured London dispersion approaches such as pairwise~\cite{Grimme-D3,TS-vdW,WannierC6} methods and 
Lifshitz-Zaremba-Kohn (LZK)\cite{Zaremba} theory. 
Unconventionally long-ranged vdW forces were observed between G and a semiconducting substrate~\cite{grafsurf},
extending well beyond the customary $\sim$10 nm scale. 
The interaction range in G was later shown to be influenced by
the dipolar response non-locality, which can substantially slow down the interaction power law decay~\cite{Dobson,Dobson-manybody,Misquitta,prb2017}.
More precisely, anisotropic nanomaterials can substantially enhance the many-body coupling between charge oscillations,~\cite{Dobson,science}
determining coherent plasma-like modes with ondulatory nature, and qualitative variations of the vdW interaction.

\begin{figure}
\includegraphics[width=5.8cm]{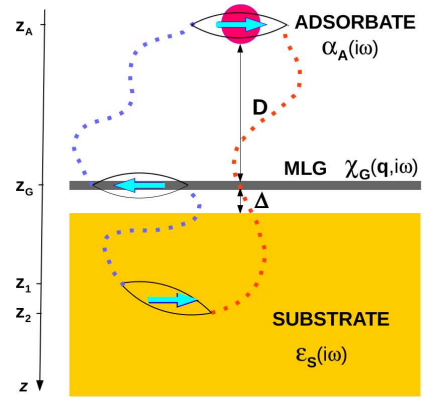}
\caption{Schematic representation of the system geometry: the adsorbate $A$ sits at distance $D$ from G, supported by a semi-infinite substrate 
at distance $\Delta$.
Light blue arrows intuitively sketch three-body dipole fluctuation modes, in correspondence of the density response functions (black closed lines).  
Dotted blue (red) wavy lines suggest attractive (repulsive) interaction between antialigned (aligned) dipoles. Clearly, while the sketched dipole moments are oriented along the G plane, all dipolar orientations will actually contribute to the interaction energy, as considered in Eq.~\eqref{energy}.}
\label{fig1}
\end{figure}

Unexpectedly, experiments~\cite{Tsoi-ACSNano} conducted by atomic force microscopy (AFM) on G-coated bulk-like SiO$_2$ further demonstrated
that G, despite single-atom thickness, can effectively screen the dispersion interaction due to the underlying substrate. The AFM tip employed to probe the surface 
was almost insensitive to the presence of SiO$_2$, so that the tip-substrate attraction was essentially determined only by G.
Moreover, this result is compatible with the very recent observation~\cite{raman2d} that a nearest neighbor effective model is sufficient to describe low-frequency (interlayer) Raman modes in stacked 2D materials.
Implications of this remarkable result may encompass stabilization of multilayer heterostructures, cleavage of 2D crystals, 
micromechanical phenomena, gas sensing or heterogeneous catalysis. 
Yet, practical application of G vdW screening  still awaits sound theoretical characterization of the underlying physical 
mechanism, as the effect could hardly be interpreted solely in terms of conventional metallic screening.

In this paper we provide theoretical rationalization of the observed G vdW {\it opacity} to weakly interacting bulk-like supports, by means of quantum many body analysis, supported by ab-initio Density Functional Theory (DFT) electrostatic calculations.
By overcoming standard local permittivity approximations to the LZK theory  we provide a
correct non-local treatment of physical adsorption of small adsorbates (atoms, molecules or nanoparticles), where both London dispersion 
(induced dipole - induced dipole coupling) and Debye forces (permanent dipole - induced dipole coupling) are considered. We find that
three-body-like repulsive terms, involving static or low-frequency dynamical polarization of both G and substrate can effectively contrast the 
attractive substrate contribution. In the same frequency regime, stark G non-locality ensures ultra-slow interaction power law decay,
mimicking the long-ranged vdW contribution due to the bulk-like underlayer.
Finally, DFT computation of electrostatic Debye interactions between a polar water molecule and SiC- and Cu(111)- supported G substrates confirms
the predicted trends, further highlighting the role of electronic structure and hybridization effects.

\section{Results and discussion}
\subsection{Response function} We conduct our calculations (details in Supplementary Material) by first deriving the fully coupled substrate response function, according to the geometry depicted in Fig.~\ref{fig1}.
Atomic units ($\hbar=e=c=m_e=1$) are adopted hereafter in order to simplify the notation.
The interacting 2D response function of G ($\chi^1_{\rm G}(q,i\omega)$) is written, as proposed in Ref.~\cite{Dobson}, in terms of the bare 
susceptibility ($\chi^0_{\rm G}(q,i\omega)$)
of the relativistic-like $\pi$ electrons with linear dispersion $\epsilon_{\pm}(\mathbf{q})=\pm v_{\rm F}|\mathbf{q}|$, coupled at the random phase approximation 
(RPA) level via the 2D Coulomb interaction $v_{\rm 2D}(q)=2\pi/q$:
\begin{equation}
\chi^1_{\rm G}(q,i\omega)=\frac{\chi^0_{\rm G}}{1-v_{\rm 2D}(q)\chi^0_{\rm G}(q,i\omega)}\,.
\end{equation}
Here $i\omega$ is the imaginary frequency, $\mathbf{q}$ the in-plane momentum, $v_{\rm F}$ the Fermi velocity, and $\chi^0_{\rm G}=-q/(4v_{\rm F}\sqrt{(1+x^2)})$, 
with $x=\omega/qv_{\rm F}$.
By assuming weak interaction between graphene and substrate we can neglect hybridization effects, and treat the two interacting response functions
as distinguishable. Accordingly, no doping in graphene is initially included in the calculation.
The interacting substrate response function $\chi_{\rm S}^1(z_1,z_2,q,i\omega)$ ($z_{1,2}$ being the coordinates orthogonal to the interface plane) 
is then coupled to $\chi^1_{\rm G}(q,i\omega)$ in the RPA fashion via the Fourier transformed interaction $v_{\rm GS}(z_1-z_2,q)=\exp(-q|z_1-z_2|)2\pi/q$.
Within a matrix formulation we can thus express the total (G+substrate) response function $\chi_{\rm T}$ as
\begin{equation}
\chi_{\rm T} = \left(
\begin{array}{cc}
\chi_{\rm G}^1 & \chi_{\rm G}^1 v_{\rm GS} \chi_{\rm S}^1  \\
\chi_{\rm S}^1 v_{\rm GS} \chi_{\rm G}^1  & \chi_{\rm S}^1  
\end{array} \right) \frac{1}{1-\chi_{\rm G}^1 v_{\rm GS} \chi_{\rm S}^1 v_{\rm GS}}\,.
\end{equation}
The total susceptibility matrix $\chi_{\rm T}$ contains the standard graphene and substrate response functions on the diagonal,
where the multiplicative term to the right-hand side corresponds to a screening renormalization.
As discussed in the following, when considering adsorption on substrate-supported graphene, the two diagonal response terms both provide attractive contributions, and screening alone could not fully justify the 
observed vdW opacity, given its equal influence on both $\chi_{\rm G}^1$ and $\chi_{\rm S}^1$.

\subsection{Adsorption energy and vdW screening} 
In order to compute the full London dispersion interaction (labeled with ${\rm L}$ hereafter) of a small-size adsorbate we will thus further include the two
off-diagonal $\chi_{\rm T}$ terms. 
The full dispersion interaction between an adsorbate with response function $\chi_{\rm A}(\mathbf{r},\mathbf{r'},i\omega)$, at large distance $D$
from the G-coated substrate can be expressed within a second order perturbative approach (corresponding to a second order expansion of the adiabatic connection fluctuation-dissipation formula) as ~\cite{Ambrosetti-JPCL,ACFDT-2013}
\begin{equation}
E^{\rm L}_{\rm vdW}=-\int_0^{\infty} \frac{d\omega}{4\pi} Tr\left[(\chi_{\rm AT}V_{\rm AT})^2\right]\,,
\label{energy}
\end{equation}
where the response function $\chi_{\rm AT}$ now comprehends both adsorbate and total substrate susceptibility $\chi_{\rm T}$, while $V_{\rm AT}$ is the
Coulomb interaction between the two (see Supplementary Material).
Making use of the above formulation one can express the total interaction energy of Eq.~\eqref{energy} as a summation
of four terms, namely $E^{\rm L}_{\rm vdW}=E^{\rm L}_{\rm I}+E^{\rm L}_{\rm II}+E^{\rm L}_{\rm III}+E^{\rm L}_{\rm IV}$, essentially corresponding to the coupling between 
the adsorbate response and the four  $\chi_{\rm T}$ components.

To proceed with our {analysis} we approximate the adsorbate response at the dipole level (polarizability $\alpha_{\rm A}$),
and express $\chi_{\rm S}^1$ in terms of the {\it average} dynamical dielectric function $\epsilon_{\rm S}(i\omega)$ of the material as proposed by
Zaremba and Kohn~\cite{Zaremba}. 
Full momentum dependence of the G response, instead
will be explicitly taken into account, given its relevance in determining the correct vdW interaction scaling law of low dimensional nanomaterials~\cite{prb2017,Dobson}.

The first term in the $E^{\rm L}_{\rm vdW}$ expansion corresponds to the expected renormalized G-adsorbate interaction, which can be recast in the following form:
\begin{equation}
E^{\rm L}_{\rm I}=\int_0^{\infty} \frac{d\omega}{2\pi} \int d^2 q \, \alpha_{\rm A}(i\omega) F_{\rm G}(i\omega,q,D,\Delta)\,,
\label{e1}
\end{equation}
where $F_{\rm G}$ (definition in the Supplementray Material)  accounts for the renormalized diagonal G response in $\chi_{\rm T}$ and for the 
G-adsorbate interaction, exponentially decaying with respect to $qD$.
Analogously, the second term derives from the renormalized bulky substrate-adsorbate interaction, and can be expressed as:
\begin{equation}
E^{\rm L}_{\rm II}=-\int_0^{\infty} \frac{d\omega}{2\pi} \int d^2 q \, \alpha_{\rm A}(i\omega) \frac{q}{2\pi} F_{\rm S}(i\omega,q,D,\Delta)\,,
\label{e2}
\end{equation}
where $F_{\rm S}$ (see Supplementary Material) now accounts for the renormalized bulky substrate diagonal response in $\chi_{\rm T}$ and for the corresponding
(exponentially decaying) substrate-adsorbate interaction.
As previously mentioned, two additional terms ($E^{\rm L}_{\rm III}$, $E^{\rm L}_{\rm IV}$) contribute to $E^{\rm L}_{\rm vdW}$, 
which derive from the off-diagonal  $\chi_{\rm T}$ terms.
These can be interpreted as renormalized three-body contributions, involving  two-fold dynamical polarization of both G and substrate, induced by
the quantum charge fluctuations in the adsorbate $A$. Intuitively, while the polarization induced over a single isolated substrate 
can be {\it antialigned} to the fluctuating dipole in $A$, thereby causing net dipole-dipole attraction, the contextual polarization of two superposed materials can result
in dipolar {\it alignment} of one of those with respect to the adsorbate $A$, thus generating a net repulsive contribution (see Fig.~\ref{fig1}).
Given the equivalence between off-diagonal terms $E^{\rm L}_{\rm III}=E^{\rm L}_{\rm IV}$, these can both be expressed as
\begin{equation}
E^{\rm L}_{\rm III}=-\int_0^{\infty} \frac{d\omega}{2\pi} \int d^2 q \, \alpha_{\rm A}(i\omega) \chi_{\rm G}^1(q,i\omega) F_{\rm S}(i\omega,q,D,\Delta)
\label{e3}
\end{equation}
Here $E^{\rm L}_{\rm II}$ and $E^{\rm L}_{\rm III}$ have opposite sign, and only differ by a single factor, namely $q/2\pi$ versus $\chi_{\rm G}^1(q,i\omega)$.
It is thus expedient to analyze the explicit expression for the G interacting response.
The low dimensionality of G and the presence of relativistic-like $\pi$ electrons  strongly influence the $\chi_{\rm G}^1$ momentum dependence.
Accordingly, at limited adsorption distances $D$ where relatively high $q$ values contribute to the integrals of Eqs.~\eqref{e2},~\eqref{e3}, 
and low imaginary frequency such that $\omega/qv_{\rm F} << 1$,  the following limiting behavior is found:
\begin{equation}
\chi_{\rm G}^1(q,i\omega) \rightarrow -\frac{q}{2\pi+4v_{\rm F}} \,.
\label{lim1}
\end{equation}
Notably, $\chi_{\rm G}^1$ exhibits the same linear dependence on momentum as the $q/2\pi$ term  in $E^{\rm L}_{\rm II}$. Contextually, the 
multiplicative factors $1/2\pi$ and $2/(2\pi+4v_{\rm F})$ (by summation of $E^{\rm L}_{\rm III}$ with $E^{\rm L}_{\rm IV}$) show comparable size, resulting in 
large cancellation of the two terms, compatibly with experimental findings~\cite{Tsoi-ACSNano}.

{\bf Limiting behaviors and dependence on adsorbate moiety:} We note that the limiting behaviour of Eq.~\eqref{lim1} is ensured under certain conditions on adsorption distance $D$ and imaginary 
frequency $i\omega$. In fact, by expressing the adsorbate polarizability in the conventional Lorentzian form $\alpha_{\rm A}(i\omega)=\alpha_{\rm A}^0/(1+\omega/\bar{\omega}_{\rm A})$,
where $\alpha_{\rm A}^0$ is the static polarizability, the condition $\omega/qv_{\rm F} << 1$ can be recast as $D<<v_{\rm F}/\bar{\omega}_{\rm A}$ (see Supplementary Material). For small oscillator frequencies of $\sim 0.05$ a.u. 
(compatible with metal nanoparticles), and renormalized~\cite{sodemann} Fermi velocity $v_{\rm F}\sim1$ a.u., this inequality 
yields $D<10$ \AA.
While the limiting $D$ value already covers  the spectrum of relevant physical adsorption distances, cancellation effects will evolve gradually with $D$, 
so that vdW {\it opacity} will persist even at larger separations, thus influencing larger scale nanoassembly phenomena.
We remark that in the above formulas frequencies are related to the Fourier components of the response function, and not to external electromagnetic fields. In fact, no external field was introduced so far, so that the whole process is governed by virtual excitations of the system.

\begin{figure}
\includegraphics[width=8.3cm]{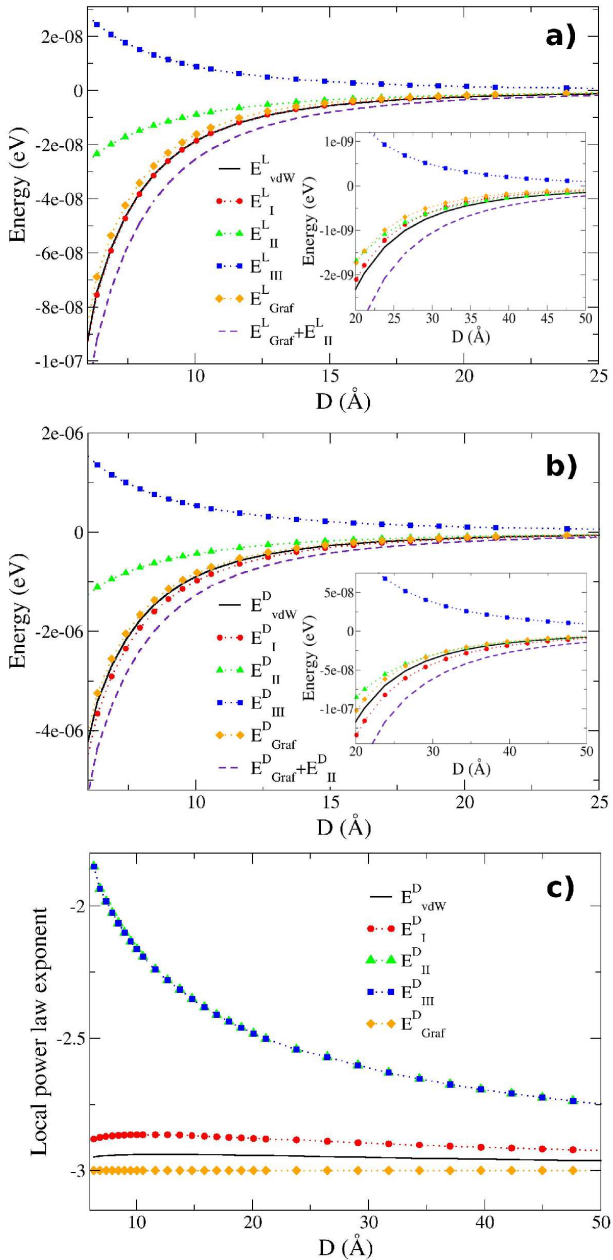}
\caption{London dispersion energy 
({\bf a)}) $E^{\rm L}_{\rm vdW}$ and corresponding energy components ($E^{\rm L}_{\rm I, II, III}$, i.e. G, substrate, and three-body contributions, respectively) for single adsorbate with $\bar{\omega}_{\rm A}=0.05$ Ha and unitary polarizability. The adsorption energy of the same adsorbate on an isolate G sheet 
($E^{\rm L}_{\rm Graf}$), 
and its sum with the substrate contribution $E^{\rm L}_{\rm II}$, are given for comparison. $\Delta$ is fixed to 3.2 \AA, compatibly with a non-covalent bonding, however, minor effects are found by reducing $\Delta$ to 1.6 \AA. In panel {\bf b)} the analysis is extended to the Debye interaction energy of a unitary dipole oriented orthogonally with respect to the surface. In panel {\bf c)} the local power law exponent is given, relative to the Debye interaction of {\bf b)}.}
\label{fig2}
\end{figure}

From Fig.~\ref{fig2} {\bf a)} we note that at low adsorbate frequency ($\bar{\omega}_{\rm A}$=0.05 Ha), 
$E^{\rm L}_{\rm vdW}$ exhibits only minor deviations from $E^{\rm L}_{\rm I}$ up to
the $\sim$25 \AA \, scale. 
The similarity persists even beyond this range, where, however $E^{\rm L}_{\rm vdW}$, $E^{\rm L}_{\rm I}$ and $E^{\rm L}_{\rm II}$ become almost indistinguishable,
while $E^{\rm L}_{\rm vdW}$ closely approaches $E^{\rm L}_{\rm II}$. 

We note that, due to strong $\chi_{\rm G}^1$ non-locality, the power law decay of $E^{\rm L}_{\rm I}$ substantially differs from standard $\sim D^{-4}$ pairwise predictions~\cite{Dobson}.  In fact, a ultra-slow $\sim D^{-3}$ scaling characterizes the short $D$ range
(due to the linear $q$ dependence of $\chi_{\rm G}^1$ at high $q$) and
gradually varies with $D$, asymptotically reaching $\sim D^{-4}$ in the large distance limit~\cite{prb2017} (well beyond the 10 nm scale).
Comparison with the dispersion interaction due to isolated G ($E^{\rm L}_{\rm Graf}$), obtained from Eq.~\eqref{e1} by setting $\epsilon_{\rm S}(i\omega)=1$,
evidences only small discrepancies with respect to  $E^{\rm L}_{\rm vdW}$.
We also underline that the large $D$ range may also be characterized by non-negligible relativistic retardation and finite temperature effects of the 
electromagnetic field, thereby possibly deviating from a naive large $D$ analysis of the above formulas. Qualitatively, the finite speed of light
is expected to damp the interaction in the large $D$ regime~\cite{Toigo,jcp18}, while finite temperatures effectively contrast retardation effects.
Overall, the relative importance of the different $E^{\rm L}_{\rm vdW}$ contributions should however be preserved, given that retardation and finite
temperature effects apply to all terms.

Depending on $\bar{\omega}_{\rm A}$ both interaction scaling laws and vdW {\it opacity}, can also vary~\cite{prb2017}, 
given the different adsorbate sensitivity to the frequency spectrum of substrate charge fluctuation modes.
As from the condition $D<<v_{\rm F}/\bar{\omega}_{\rm A}$ for Eq.~\eqref{lim1}, we note that adsorbates with high $\bar{\omega}_{\rm A}$ may exhibit sizeable deviations from the {\it strongly opaque}
regime at shorter $D$. In fact, vdW {\it opacity} is most effective in the low frequency limit, and could thus be experimentally 
tuned by an appropriate choice of the adsorbed moiety. 
We remark that the oscillator frequency is physically related to the adsorbate HOMO-LUMO (highest occupied-lowest unoccupied molecular orbital) gap, and effectively accounts for the confinement of the electronic cloud. For instance, atomic Na is characterized by $\bar{\omega}_{\rm A} \sim 0.08$ Ha, while the oscillator frequency is roughly ten times larger for Ar monomers. Low oscillator frequencies (and correspondingly high screening) are also to be expected in metal nanoparticles due to high charge mobility, compatibly with high vdW {\it opacity} observed via AFM metallic tip.

\subsection{Debye electrostatic interactions} Interestingly, vdW {\it opacity} is not restricted to London dispersion forces only, but also extends to Debye interactions,
arising in the presence of polar adsorbates. This is understood given the evident analogies between $E^{\rm L}_{\rm vdW}$ 
 and the Debye interaction $E^{\rm D}_{\rm vdW}$ 
\begin{equation}
\label{edebye}
E^{\rm D}_{\rm vdW}=-\frac{1}{2}Tr\left[(M_{\rm AT}V_{\rm AT})^2\right]\arrowvert_{(i\omega=0)}\,,
\end{equation}
where $M_{{\rm AT}, ij}=-\delta_{ij}\delta_{i1}\partial_{z_A} d^2_{\rm A}\partial_{z_A}+(1-\delta_{ij}\delta_{i1})\chi_{{\rm AT},ij}$. The absence of frequency integration 
in $E^{\rm D}_{\rm vdW}$ (cfr. Eq.~\eqref{energy}) ensures dependence on the
static ($i\omega=0$) response only, while the permanent dipole moment $d_{\rm A}$ (here assumed orthogonal to the G plane for simplicity) ultimately replaces 
$\alpha_{\rm A}$, inducing charge polarization in G and substrate. We also stress that the Debye interaction energy scales roughly quadratically with respect to the adsorbate dipole moment, as from Eq.~\eqref{edebye}. Screening effects in the adsorption of different polar moieties on supported G can thus be inferred from the present analysis by suitable energy rescaling.
In Fig.~\ref{fig2} {\bf b)} we report an energetic analysis for the Debye interaction, in analogy with London dispersion. All terms labeled with
$D$ provide straightforward extensions of the above London terms to the electrostatic Debye case. Qualitative similarities between
panels {\bf a)} and {\bf b)} in Fig.~\ref{fig2} are self evident, while the static nature of Debye forces implies slightly longer-ranged vdW opacity.
By a {\it local power law} analysis (see Fig.~\ref{fig2} {\bf c)}), we can access the exponents determining the 
scaling of the Debye interaction in terms of the adsorption distance (given as the slope in logarithmic scale of the energy with respect to $D$), 
in analogy with Refs.~\cite{science, prb2017, Dobson}.
Although $i\omega=0$ automatically implies linear $q$ dependence of $\chi_{\rm G}^1$, resulting in constant $~D^{-3}$ scaling of $E^{D}_{\rm Graf}$ (at variance with London dispersion),
residual power law variations are still found for $E^{\rm D}_{\rm I,II,III}$. These are due to the overall renormalization factor, which acquires larger
relevance at low $q$. Remarkably,  $E^{\rm D}_{\rm vdW}$ also exhibits very similar
local power law scalings with respect to both $E^{\rm D}_{Graf}$ and  $E^{\rm D}_{\rm I}$, hence
confirming once more the effectiveness of static vdW {\it opacity}.

\subsection{Ab-initio modeling of realistic substrates} The results presented so far rely on approximate $\pi$-electron G susceptibility (neglecting $\sigma$ orbitals), and do
not include possible G-substrate hybridization effects. In order to test the reliability of our predictions and better assess both
validity and limits of vdW {\it opacity} in realistic materials, we thus complement our study with ab-initio DFT calculations, within the
semi-local Perdew Burke Ernzerhof~\cite{PBE} approximation, through Quantum Espresso~\cite{QE} simulation package. While semi-local DFT does not correctly capture long-ranged dispersion correlations, it actually provides a reliable description of many-body electrostatic effects, arising for instance upon adsorption of polar molecules on arbitrary substrates. In fact, at variance with London dispersion, the Debye interaction is not a long-range correlation effect, and emerges instead as a response to the {\it external potential} induced by the adsorbate dipole.
Moreover,  the DFT electrostatic response can naturally account for all G electrons, further including G-substrate 
hybridization effects. 
In order to resolve the relevant momentum dependence of the G density response we adopted a dense K-point sampling of the 2D Brillouin zone (24x24 regular mesh -- 
computational details are reported in the Supplementary Material). Wide vacuum spacing along $z$ (minimum 30 \AA\, from the adsorbate 
to the closest substrate image) was also introduced in order to minimize the interaction with periodic replicas. 

\begin{figure}
\includegraphics[width=8.5cm]{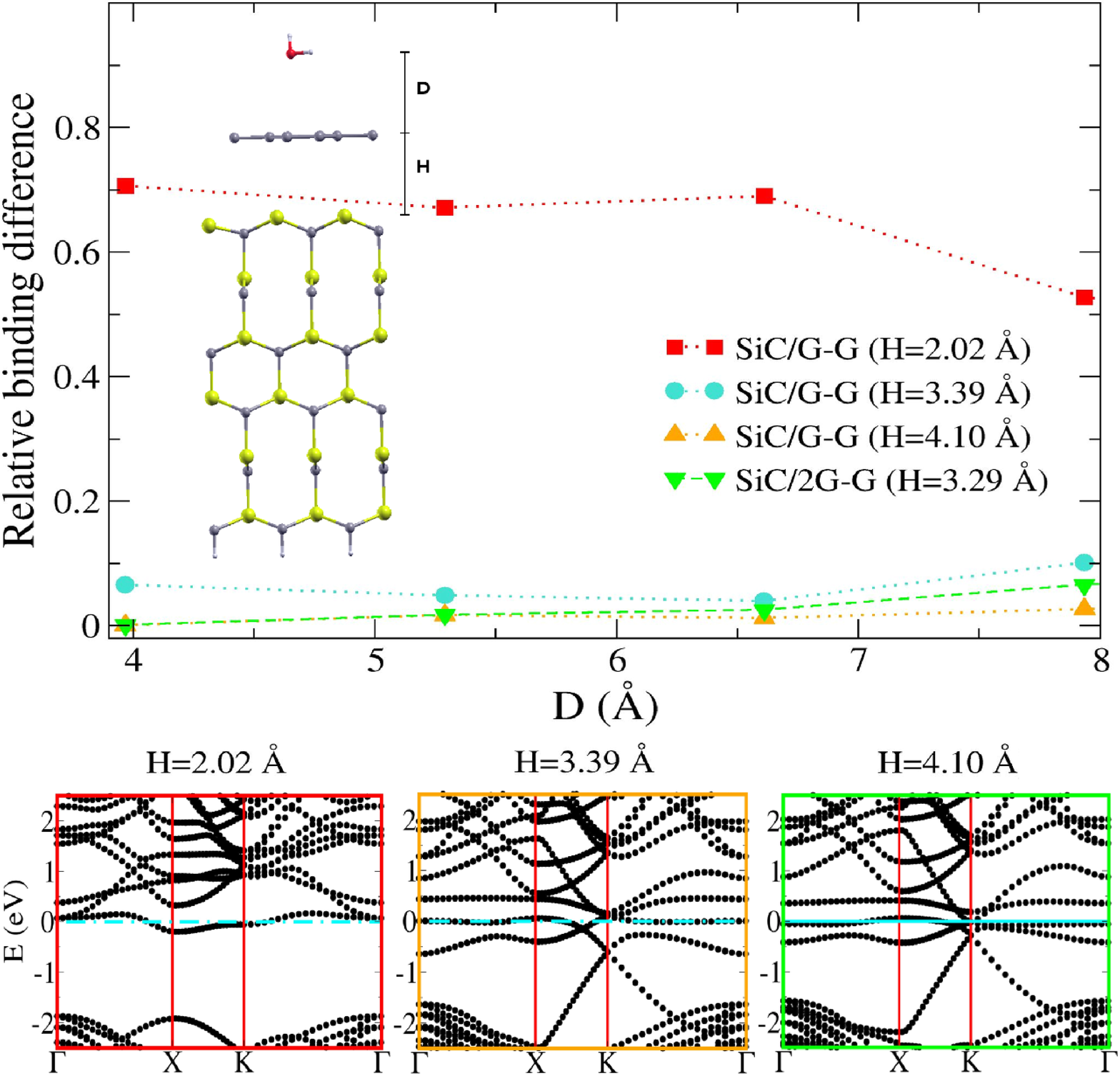}
\caption{
Difference (in modulus) between binding energies for a single water molecule adsorbed on SiC/G and on free-standing G, divided by the binding energy for adsorption on SiC/G, at different 
SiC-G separations $H$. Analogous relative binding energy differences are presented also for G on G-coated SiC (Sic/2G).
Inset illustrates the simulation cell geometry. Lower panels report the electronic bands of G/SiC at the considered separations $H$  (Fermi levels are indicated by light blue dash-dotted lines). At larger $H$, where hybridization effects are gradually reduced, vdW (Debye) {\it opacity} becomes 
evidently more pronounced.}
\label{fig3}
\end{figure}

We conduct our analysis by first studying the {\it opacity} of a G buffer layer on Silicon Carbide. As from Fig.~\ref{fig3}, when 
the G buffer layer sits at equilibrium distance from SiC (2.02 \AA), strong electronic hybridization occurs due to chemical 
bondings. We note that the Fermi level intersects a weakly dispersive band, while a large gap dictated by the SiC electronic structure exists below this band, at variance with pristine G.
The conic band dispersion of free-standing G is strongly altered, and G electrostatic screening is poor. Accordingly, the adsorption energy of a 
H$_2$O molecule adsorbed on G/SiC shows sizeable deviation from that of H$_2$O on free-standing G. We underline that an arbitrary orientation of H$_2$O was selected, with non-zero electric dipole components both parallel and orthogonal to the G plane, in order to ensure maximum generality. By gradually lifting the G 
buffer from SiC, instead, hybridization is suppressed and screening is visibly enhanced, so that quasi complete screening 
is achieved at G-SiC separation of $\sim$4 \AA \,(as a comparison to Fig~\ref{fig3}, the estimated H$_2$O-G binding amounts to $\sim$ 8$\cdot$20 meV at $D$=3.97 \AA). This can be directly inferred from the data reported in Fig.~\ref{fig3} at fixed G-H$_2$O distance, and variable G-substrate separation. 
For completeness, we also note that the estimated interaction energy between H$_2$O (at the given orientation) and SiC at the considered adsorption distances $H+D$, where $H$=3.97 \AA, remains roughly one order of magnitude larger then the binding energy difference between H$_2$O on G and H$_2$O on G+SiC, in spite of the large $H+D$ distance. 

While lifting the G buffer might be experimentally challenging, the growth of G on G-buffer-coated SiC is routinely 
accomplished~\cite{sic-graf,sic-a} via thermal decomposition techniques. Notably, also in this case, the topmost G layer (at the equilibrium distance of $\sim$3.29 \AA \, from the 
G buffer) exhibits minor hybridization effects, and strong vdW {\it opacity} is again evident from Fig.~\ref{fig3}.

\begin{figure}
\includegraphics[width=8.5cm]{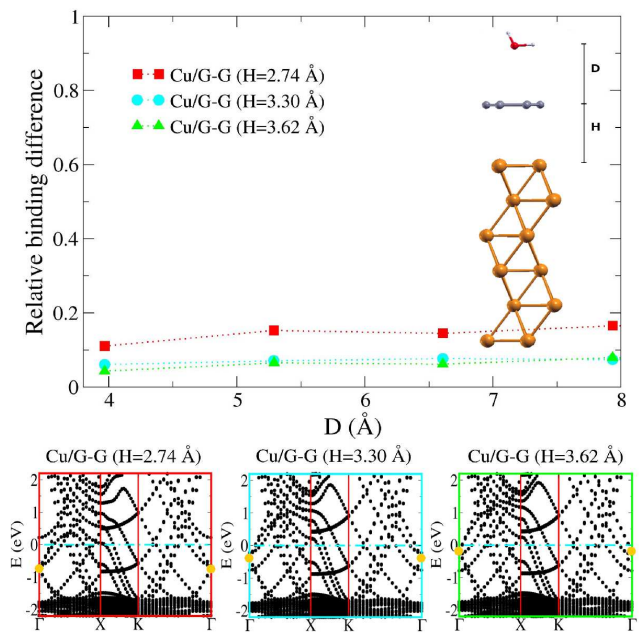}
\caption{
Difference (in modulus) between binding energies for a single water molecule adsorbed on Cu(111)/G and on free-standing G, divided by the binding energy for adsorption on Cu/G, for different values
of the Cu-G separation $H$. Inset illustrates the simulation cell geometry. Even in the presence of a metallic substrate, the  trend observed in SiC/G screening is preserved. Lower panels show the band structure of the above G/Cu structures (Fermi levels are indicated by light blue dash-dotted lines). Due to the adopted supercell
symmetry, the G Dirac cone (located at K when considering the G primitive cell) is shifted to the supercell $\Gamma$ point. Yellow dots highlight the position of G Dirac points.}
\label{fig4}
\end{figure}

Given the growing relevance of transition metal surfaces for large scale fabrication of high quality G layers via chemical vapor deposition~\cite{Batzill2014}, we finally considered Cu(111) as a substrate. As from
Fig.~\ref{fig4}, Cu(111) causes a n-doping of G~\cite{Walter}, shifting the Fermi energy above the Dirac point (located here at $\Gamma$ due to the adopted supercell).
While the Fermi energy shift is sizeable below the Cu-G equilibrium distance~\cite{Batzill2014} (i.e. $H$=3.30 \AA), by lifting the G layer the Fermi level
gradually aligns with the Dirac point, due to weaker G-Cu density overlap. Also in this case vdW screening becomes most effective at larger $H$,
and high {\it opacity} is recovered close to Fermi level-Dirac point alignment, consistently with experimental observations\cite{Tsoi-ACSNano}.

The present results are expected to extend to further 2D materials exhibiting Dirac cone dispersion, such as Silicene, Germanene and Stanene~\cite{silicene,stanene}, 
given the evident analogies in geometry and electronic structure, and the generality of our theoretical approach. We remark that the strong screening predicted for monolayer materials substantially depends on the non-locality of the density response (exhibiting quasi-linearity with respect to momentum). By comparison, thin dielectric films exhibiting negligible momentum dependence of the permittivity would deviate from the present predictions, leading to {\it conventional} dielectric screening. Moreover, three-body cancellation effects are expected to vanish at large adsorption distance D for {\it conventional} finite-thickness films: in fact, the layer-adsorbate energy contribution falls off as $D^{-4}$, at variance with the substrate-adsorbate term, which scales as $D^{-3}$.
While above calculations specifically rely on Dirac cone dispersion, we recall that the key ingredient ensuring the correct scaling of the interactions and the many-body cancellation effects is the strong non-locality of the G dipolar response (essentially determined by its momentum dependence). Previous calculations conducted on quasi 2D MoS$_2$ ~\cite{science,prb2017} have equally evidenced strongly non-local dipolar response, and close analogies
to the vdW scalings of graphene. The combination of quasi 2D geometry and response non-locality thus indicates extendibility of the present
theory to 2D transition metal dichalcogenides~\cite{dichalcogenides}, coherently with the experimental evidence~\cite{Tsoi-ACSNano}.
On the other hand, the observed reduction of vdW {\it opacity} upon doping of G can be rationalized in terms of modified electronic structure, and loss of non-locality.
In fact, dopants can effectively vary the Fermi level~\cite{Graphene-doping}, {\it hiding} or disrupting the peculiar Dirac cone structure, in analogy with G-substrate hybridization. Moreover, they can contextually introduce 
effective inhomogeneities 
in G, with a consequent loss of translational invariance that can destructively interfere with collective charge displacement modes.
Finally, we comment on the role of $v_{\rm F}$ in graphene: at $v_{\rm F}\rightarrow 0$, when Dirac cones tend to {\it flatten}, one finds that the linear $q$-dependence of $\chi_{\rm G}^1$ given by Eq.~\eqref{lim1} is lost, and a quadratic dependence in $q$ is found instead even at large momenta. Once again, the loss of non-locality in the dipolar response implies a weakening of the vdW {\it opacity}.

\section{Conclusions}
In conclusion, we provided theoretical confirmation and rationalization of the experimentally observed vdW opacity, caused by G on weakly interacting substrates 
upon physical adsorption. 
A repulsive three-body-like contribution contrasts the expected attractive term deriving from the underlying substrate, so that the total vdW adsorption 
energy is well approximated by the sole G-adsorbate contribution. Stark non-locality of the G electron density response plays a main role 
in this mechanism, enforcing the correct interaction scaling law, and ensuring effective compensation. The phenomenon is most effective in the
low frequency regime, and extends beyond London dispersion, encompassing electrostatic Debye forces. 
Owing to the present insights and the combined theoretical/experimental validation, G and related quasi 2D materials 
emerge as a promising and accessible tool for {\it filtering}, modulating or effectively switching vdW interactions at interfaces, thus opening novel perspectives
for detailed experimental control of heterostructures, surface phenomena and assembly of complex nanomaterials.

\section{acknowledgement}
We acknowledge insightful discussion with A. Tkatchenko, R. DiStasio Jr., and F. Toigo. We also acknowledge financial support from Cassa di Risparmio di Padova e Rovigo (Cariparo)--grant EngvdW. Computational resources were granted by CINECA.

\bibliography{literature}

\end{document}